\documentclass[twocolumn,10pt]{asme2e}

\usepackage{graphicx}
\usepackage{hyperref}
\hypersetup{
    colorlinks=true,
    linkcolor=blue,
    citecolor=blue,
    urlcolor=blue,
}
\usepackage{float}
\usepackage{svg}
\usepackage{amsmath}
\usepackage{subfigure}

\confshortname{IMECE 2021}
\conffullname{the ASME 2021 \\ The International Mechanical Engineering Congress and Exposition}

\confdate{1-5}
\confmonth{November}
\confyear{2021}
\confcity{Virtual Conference, Online}
\confcountry{USA}

\papernum{IMECE2021/69657}

\title{Neural Differential Equations for Inverse Modeling in \\ Model Combustors}

\author{Xingyu Su                             
    \affiliation{
	Tsinghua University \\
	Beijing 100084, China \\
    su-xy19@mails.tsinghua.edu.cn
    }
}

\author{Weiqi Ji
    \affiliation{
	Massachusetts Institute of Technology \\
	Cambridge, MA 02139 \\
    weiqiji@mit.edu
    }
}

\author{Long Zhang
    \affiliation{
	Tsinghua University\\
	Beijing 100084, China\\
	long-zha19@mails.tsinghua.edu.cn
    }
}

\author{Wantong Wu                             
    \affiliation{
	Tsinghua University \\
	Beijing 100084, China \\
    wwt15@mails.tsinghua.edu.cn
    }
}

\author{Zhuyin Ren \thanks{Address all correspondence to this author.}                           
    \affiliation{
	Tsinghua University \\
	Beijing 100084, China \\
    zhuyinren@mail.tsinghua.edu.cn
    }
}

\author{Sili Deng \thanks{Address all correspondence to this author.}
    \affiliation{
	Massachusetts Institute of Technology \\
	Cambridge, MA 02139 \\
    silideng@mit.edu
    }
}

\begin{document}
\maketitle

\begin{abstract}
{\it
Monitoring the dynamics processes in combustors is crucial for safe and efficient operations. However, in practice, only limited data can be obtained due to limitations in the measurable quantities, visualization window, and temporal resolution. This work proposes an approach based on neural differential equations to approximate the unknown quantities from available sparse measurements. The approach tackles the challenges of nonlinearity and the curse of dimensionality in inverse modeling by representing the dynamic signal using neural network models. In addition, we augment physical models for combustion with neural differential equations to enable learning from sparse measurements. We demonstrated the inverse modeling approach in a model combustor system by simulating the oscillation of an industrial combustor with a perfectly stirred reactor. Given the sparse measurements of the temperature inside the combustor, upstream fluctuations in compositions and/or flow rates can be inferred. Various types of fluctuations in the upstream, as well as the responses in the combustor, were synthesized to train and validate the algorithm. The results demonstrated that the approach can efficiently and accurately infer the dynamics of the unknown inlet boundary conditions, even without assuming the types of fluctuations. Those demonstrations shall open a lot of opportunities in utilizing neural differential equations for fault diagnostics and model-based dynamic control of industrial power systems.
}


\end{abstract}

\section*{1. INTRODUCTION} \label{sec:intro}

Oscillation in combustors can degrade the efficiency and safety of the combustion system. Unsteadiness in the upstream is one of the major causes of the oscillation, and thus monitoring the upstream conditions is important for control and operation. However, in practice, it is difficult to measure the upstream conditions due to the challenges in diagnostic at harass environment and the constraints of cost. Therefore, computational tools aiming at the inference of unknown information based on available measurements have been developed and are often termed as inverse models \cite{cui2018inverse, perakis2019heat, lee2011inverse}. Conventional inverse modeling methods are usually limited to a few unknown parameters and suffering from the curse of dimensionality since the computational cost increases significantly as the number of unknown quantities increases.

In the present work, we propose to use neural networks for approximating the high dimensional unknown quantities and train the neural networks efficiently with available measurements by differential programming. In order to achieve good extrapolation capability outside of the training dataset, supervised learning methods usually require a large number of training datasets that can cover a wide range of conditions. For the neural differential equation, physical constraints are imposed during the training by augmenting the neural network models with the governing differential equations, such that the training can be viewed as self-supervised learning to avoid the need for obtaining labeled training data, in this case, the unmeasurable, which are intrinsic obstacles for modeling industrial combustors with data-driven approaches.

We demonstrated the inverse modeling approach in a model combustor system by simulating oscillating combustion in an industrial hot air heater with a perfectly stirred reactor (PSR). Industrial combustion devices such as hot air heaters are widely used for burning non-standard low-calorific value (NLCV) gases, such as blast furnace gas and coke oven gas, which are prone to combustion instability. Zhang et al. \cite{zhang2020oscillation} studied the oscillating combustion in reactors resulting from periodic fuel burnt out, fuel replenishment, and ignition governed by the subtle mixing-chemical kinetics interaction or due to the inlet flow rate fluctuation under industrial conditions. Zhang et al. utilized forward modeling to investigate the oscillation of combustion via unsteady PSR simulation. Given the initial conditions and boundary conditions, one can integrate the governing ordinary differential equations (ODEs) to get the temporal evolution profile of temperature and species concentrations. However, in practical industrial scenarios, it is often difficult to measure the boundary conditions with satisfying accuracy, especially for species concentrations. The experimental data that can be effectively obtained are the temperature and flow velocity at the combustor outlet. In this study, we try to infer the inlet fluctuations in compositions and/or flow rate given the sparse measurements of the temperature at the combustor outlet. The fluctuation in the inlet conditions is usually time-dependent, such that one needs to use a large number of data points to resolve the evolution of the inlet conditions, and thus the inverse modeling involves high dimensional unknown parameters.

This work aims to tackle the challenges in achieving scalability of the data-driven inverse modeling of unknown measurements in model combustors using techniques developed for training deep neural network models. Specifically, we first parameterize the fluctuations using a neural network model. We then employ the stochastic gradient descent (SGD) optimizer to learn the inlet boundary conditions in an unsteady PSR. The SGD optimizer is widely applied in the deep learning communities for its efficiency as well as scalability in dealing with large datasets, and its robustness in optimizing high-dimensional non-convex neural network models. Our recent work \cite{ji2021autonomous} has also shown that a chemical reaction network is equivalent to a neural network with a single hidden layer. Similarly, solving ordinary differential equations (ODEs) of reaction network models is equivalent to solving infinite-depth deep residual networks \cite{chen2018neural}, which further rationalizes exploiting SGD in training ODEs.

However, to exploit neural networks in combustion simulation and optimization, one needs a well-established software ecosystem that can efficiently and accurately compute the gradient of simulation output to model parameters. For instance, the finite difference method (often termed as the "brute-force method") usually suffers both the computational inefficiency due to the cost increases with the number of parameters and the numerical inaccuracy due to the truncation error \cite{baydin2018automatic}. Conversely, gradient evaluation methods based on auto-differentiation (AD) have shown both efficiency and accuracy in the training of large-scale deep neural network models in a variety of tasks. Many open-source AD packages have been developed in the last decade, including TensorFlow \cite{abadi2016tensorflow}, Jax and PyTorch \cite{paszke2019pytorch} in Python, ForwardDiff.jl and Zygote.jl \cite{revels2016forward} in Julia. Thus, this work utilized a recently developed differentiable combustion simulation package Arrhenius.jl \cite{jiarrheniusgithub} to enable SGD for combustion model optimizations and applies Arrhenius.jl to infer the boundary conditions of the model combustor described by PSR. Specifically, Arrhenius.jl, the differential programming ecosystem in Julia, was employed to encode and solve the governing ODEs for a perfectly stirred reactor, conduct auto-differentiation, and train the neural network models. Various types of fluctuations in the upstream, as well as the responses at the combustor's outlet, were synthesized to train and validate the algorithm. The results demonstrated that the approach can efficiently and accurately infer the dynamics of the inlet flow and calorific fluctuations.

This paper is structured as follows: we first introduce the package of Arrhenius.jl, the governing equations for the combustor, and the inverse modeling using neural differential equations; we then demonstrate the approach in inverse modeling during two representative scenarios: fuel switching and inflow fluctuations. Finally, we conclude and discuss the opportunities and challenges of employing the current approach for the dynamic control of industrial power systems.

\section*{2. METHODS} \label{sec:methods}

\subsection*{2.1 Arrhenius.jl}

Arrhenius.jl is built using the programming language of Julia to leverage the rich ecosystems of auto-differentiation and differential equation solvers. Arrhenius.jl does two types of differentiable programming: (i) it can differentiate elemental computational blocks. For example, it can differentiate the reaction source term with respect to the kinetic and thermodynamic parameters as well as species concentrations; (ii) it can differentiate the entire simulator in various ways, such as solving the adjoint sensitivities \cite{rackauckas2017differentialequations}. The first type of differentiation plays the basis role for the second type of high-level differentiation. Arrhenius.jl offers the core functionality of chemical reaction simulations in native Julia programming, which allows users to readily build the applications on top of Arrhenius.jl and exploit various approaches to do high-level optimizations.

\begin{figure*}
    \centering
    \includegraphics[width=0.8\textwidth]{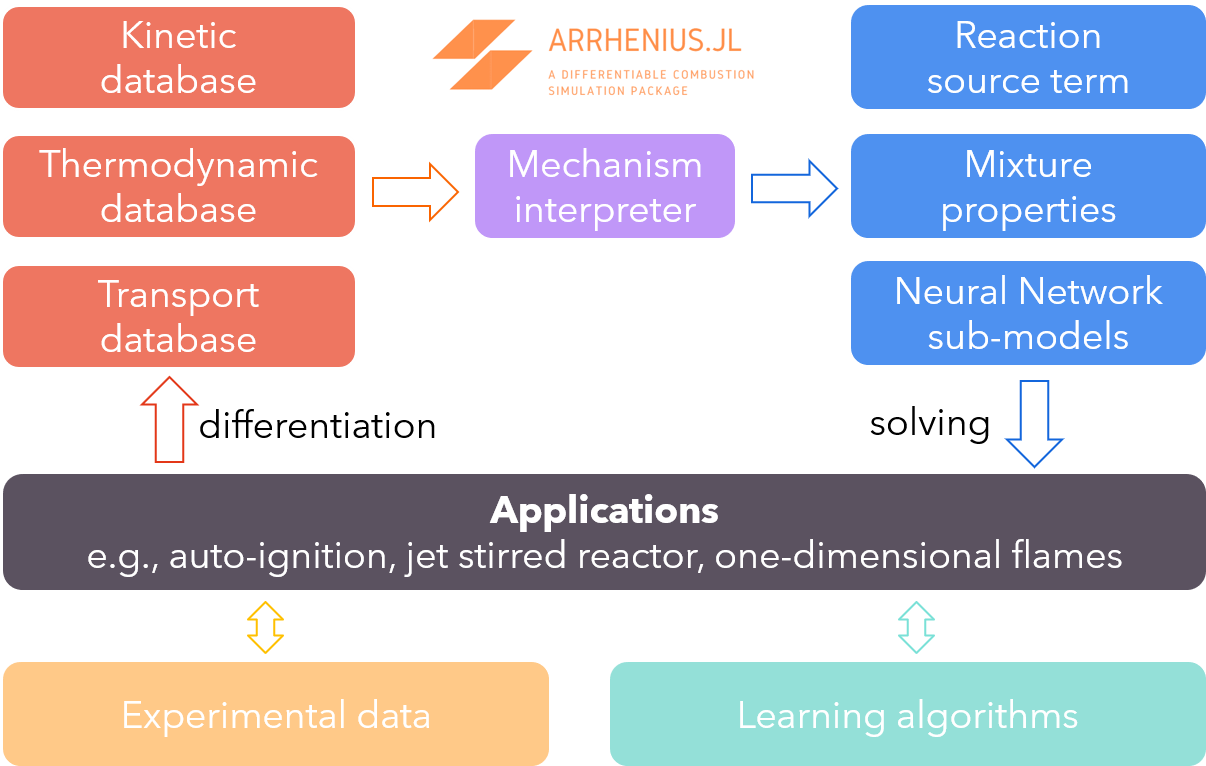}
    \caption{Schematic diagram showing the structure of the Arrhenius.jl \cite{jiarrheniusgithub} package.}
    \label{fig:schem}
\end{figure*}

Figure \ref{fig:schem} shows a schematic diagram of the structure of the Arrhenius.jl package. The package reads in the reaction mechanism files in YAML format as the same as those in the Cantera community, which contains the kinetic models, thermodynamic, and transport databases. The core functionality of Arrhenius.jl is to compute the reaction source terms and mixture properties, such as heat capacities, enthalpies, entropies, Gibbs free energies, etc. Furthermore, Arrhenius.jl makes it possible for users to define neural network models as submodels and augment them with existing physical models. For example, one can use a neural network submodel to represent the unknown chemical reaction pathways and exploit various scientific machine learning methods to train the neural network models, such as neural ordinary differential equations \cite{chen2018neural, kim2021stiff} and physics-informed neural network models \cite{raissi2019physics, ji2020stiff}. One can then implement the governing equations for different applications with these core functionalities and solve the governing equations using classical numerical methods or neural-network-based solvers. Arrhenius.jl provides solvers for canonical combustion problems, such as simulating the auto-ignition in constant volume/pressure reactors and oxidation in perfectly stirred reactors.

Compared to the legacy combustion simulation packages, Arrhenius.jl can not only provide predictions given the physical models but also the capability of optimizing model parameters or boundary conditions given experimental measurements. By efficiently and accurately evaluating the gradient of the solution outputs to the model parameters, experimental data can be incorporated into the simulation pipeline to enable data-driven modeling with deep learning algorithms.

\subsection*{2.2 Governing Equations for the Model Combustor}

Following the work of \cite{zhang2020oscillation, zhang2021ann}, we employ a non-adiabatic unsteady PSR to reveal the complex dynamics exhibited in industrial hot air heaters. The chemical kinetic model employed is DRM22, a reduced version of GRIMECH 1.2 obtained by Kazakov et al. \cite{kazakov1994drm22}, which consists of 22 species and 104 reactions.

The hot air heater in this study operates under constant atmospheric pressure. The heater chamber is cylindrical with a constant volume of 13.82 $m^3$ as detailed in \cite{zhang2020oscillation}. The mass flow rate of the inlet mixture $m_a$ is 2.75 $kg/s$, and the residence time $\tau_{res}$ is calculated according to Eq. \ref{eq:tau_res},
\begin{equation}
    \tau_{res} = \frac{\rho V}{m_a},
    \label{eq:tau_res}
\end{equation}
where $\rho$ is the density of the mixture in the reactor and $V$ is the volume of the hot air heater. With the assumption of perfect mixing in the reactor, the dynamics inside the combustor can be modeled by a non-adiabatic PSR governed by the following set of ODEs,
\begin{align}
    \frac{d \pmb{Y}}{dt} &= \frac{\pmb{Y}_{in}-\pmb{Y}}{\tau_{res}} + \pmb{S}(\pmb{Y}),
    \label{eq:dYdt}
    \\
    \frac{dT}{dt} &= \frac{\sum_i q_i R_i}{\rho C_p} - \frac{T-T_{in}}{\tau_{res}} - \frac{Q_{loss}(T-T_a)}{\rho C_p},
    \label{eq:dTdt}
\end{align}
where $\pmb{Y}_{in}$ and $T_{in}$ are the species mass fraction and temperature of the inlet stream, $T_a$ is the ambient temperature, $Q_{loss}$ is the heat loss coeﬃcient, $T $ is the mixture temperature inside the chamber,  $C_p$ is the specific heat of the mixture inside the reactor, $R_i$ is the $i-th$ reaction rate, $q_i$ is the exothermicity of the $i-th$ reaction, and the vector $\pmb{S}(\pmb{Y})$ refers to the chemical reaction source terms of the compositions. The combustion characteristics of PSR are determined by the boundary conditions of $\tau_{res}$, $\pmb{Y}_{in}$, $T_{in}$, $P$, $T_a$ and $Q_{loss}$ and the initial conditions.

This work considers two kinds of NLCV gases, the coke oven gas and the blast furnace gas. For both two kinds of fuels, the hot air heater is operated under the following conditions: $P = 101325 Pa$, $T_{in} = 300 K$, $T_a = 873 K$, $Q_{loss} =4 J K^{-1} m^{-3}s^{-1}$, and $\tau_{res} =1.0 s$. The heat loss coeﬃcient $Q_loss$ is estimated by calculating the equivalent thermal resistance of the heater wall. Typical compositions of the NCLV gases applied in this study are listed in Table \ref{tab:composition}. The initial conditions of the PSR are the adiabatic and isobaric chemical-equilibrium state of the inlet mixture. Equations (\ref{eq:dYdt}-\ref{eq:dTdt}) are numerically integrated with the ODE solver $TRBDF2()$ in the Julia package DifferentialEquations.jl \cite{rackauckas2017differentialequations}.

\begin{table}[htbp]
  \caption{Typical fuel compositions of coke oven gas (COG) and blast furnace gas (BFG) used (volume fraction) \cite{zhang2020oscillation}}
  \label{tab:composition}
  \begin{center}
    \begin{tabular}{l l l l l l}
      \hline
      Fuel composition  &      &   H2   &   CO   &   CH4  &   N2   \\
      \hline
      Coke oven gas     & COG1 & 50.0\% &  9.0\% & 30.0\% & 11.0\% \\
                        & COG2 & 60.0\% &  9.0\% & 25.0\% &  6.0\% \\
      Blast furnace gas &  BFG &  2.5\% & 22.5\% &  0.0\% & 75.0\% \\
      \hline
    \end{tabular}
  \end{center}
\end{table}

\subsection*{2.3 Inverse Modeling}

\begin{figure*}
    \centering
    \includegraphics[width=0.9\textwidth]{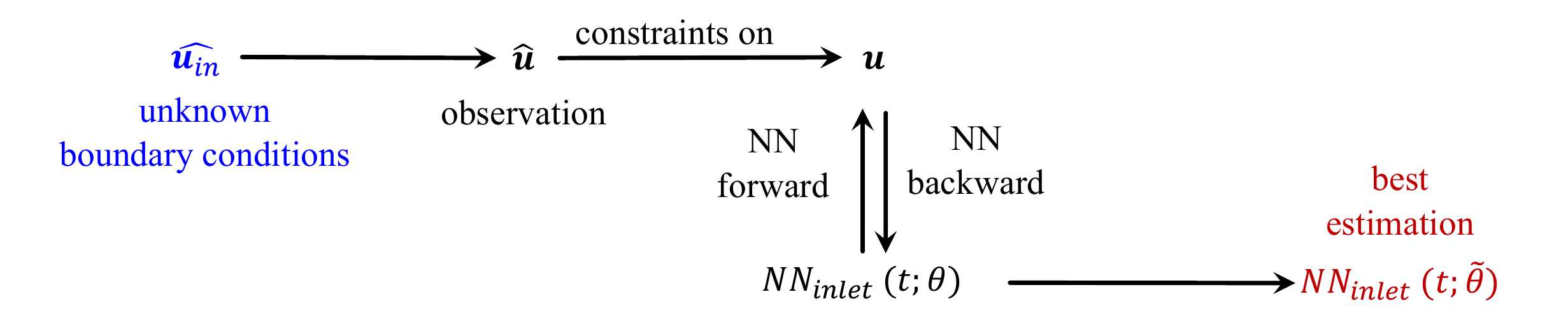}
    \caption{Schematic diagram of the inference of boundary conditions via neural networks.}
    \label{fig:psr_schem}
\end{figure*}

As discussed in the introduction, inverse modeling mainly involves unknown inlet conditions. To alleviate the high dimensionality of the time-dependent inlets, we use a neural network model to represent the inlet conditions. The neural network takes the input of time and outputs the mixture compositions and mass flow rate of the inlet stream. Recent work of physics-informed neural network \cite{raissi2019physics} has demonstrated that the neural network representation of time-dependent state variable not only tackles the curse of dimensionality but also has the potential to learn the physics information in the training data and predict the state variable beyond the training time domain. Therefore, this work employs a neural network model to infer the unknown inlet conditions, as outlined in Fig. \ref{fig:psr_schem}. The true inlet conditions $\hat{\pmb{u}_{in}}$ are essentially unknown, and the optimization process will make the state variables $\pmb{u}$ fit with the observations $\hat{\pmb{u}}$, and thus we get the $NN_{inlet}(t;\tilde\theta)$ as the best estimation of the boundary conditions.

Without loss of generality, we can rewrite Eqs. \ref{eq:dYdt} and \ref{eq:dTdt} as
\begin{equation} \label{eq:dudt}
    \frac{d\pmb{u}}{dt} = f(\pmb{u}, \pmb{u}_{in}),
\end{equation}
where $f$ refers to the right hand side of the ODEs. $\pmb{u}$ refers to the vector of state variables $[\pmb{Y}, T]$. We then use a neural network model $NN_{inlet}(t;\theta)$ to model the inlet mixture, i.e.,
\begin{equation} \label{eq:nn}
    \pmb{u}_{in} = NN_{inlet} (t; \theta),
\end{equation}
where $\theta$ refers to all of the weights and bias of the neural network model.
\begin{equation} \label{eq:dudt_nn}
    \frac{d\pmb{u}}{dt} = f(\pmb{u}, NN_{inlet}(t; \theta)).
\end{equation}

For a given initial condition $\pmb{u}_0$, the ODEs of Eq. \ref{eq:dudt_nn} can be solved with a stiff ODE solver from DifferentialEquations.jl and results in the following solution array:
\begin{equation} \label{eq:solution}
    \pmb{u}(t) = \text{ODESolve}(f, NN_{inlet}, \pmb{u}_0, t; \theta).
\end{equation}
We can then compare the predicted evolution of compositions inside the combustor with available measurements and guide the optimization of the neural network model. Specifically, with observations $\pmb{\hat{u}}$ as the training data, one can define the loss function as the mean square error (MSE) between the solution array and the training data via
\begin{equation} \label{eq:loss}
    Loss = MSE\left(\pmb{u}, \pmb{\hat{u}}\right).
\end{equation}

We employ the algorithm of ForwardDiffSensitivity to compute the gradient of the loss function to the neural network weights. ForwardDiffSensitivity is an implementation of the discrete forward sensitivity analysis through ForwardDiff.jl \cite{revels2016forward}. Note that accurate gradient computation is crucial for training neural network models as the optimization problem is usually stiff and it is difficult to reach a good minimum with noisy gradient.

With the computed gradient, we then employ the SGD optimizer to optimize the neural network models. SGD optimizer is usually very sensitive to the hyper-parameters, such as the learning rate, initialization, etc. We will present the details of the optimizer in the next section. We then treat the neural network that minimizes the mismatch between the predicted and measured composition evolution inside the combustor as the best estimation of the inlet conditions.

\section*{3. RESULTS} \label{sec:results}

Non-steady inlet boundary conditions are common causes for the oscillation of temperature and species concentrations inside the combustor. Industrial operations using the NLCV fuels are mainly concerned with two types of combustion instabilities. One is due to fuel switching and the other is due to the flow rate and calorific fluctuations in the inlet flow. In this section, we demonstrate the present method in the two representative cases with industry relevant amplitudes and frequencies. We also design different initialization method and different noise levels to demonstrate the robustness of the optimization process. We will first demonstrate the identification of fuel switching from the measurement of temperature history and then demonstrate the inference of inlet flow rate and calorific fluctuations from the measurement of temperature and CO profiles.

\subsection*{3.1 Inverse Modeling for Fuel Switching}

\begin{figure*}
    \centering
    \subfigure[inlet conditions]{
        \includegraphics[width=0.34\textwidth]{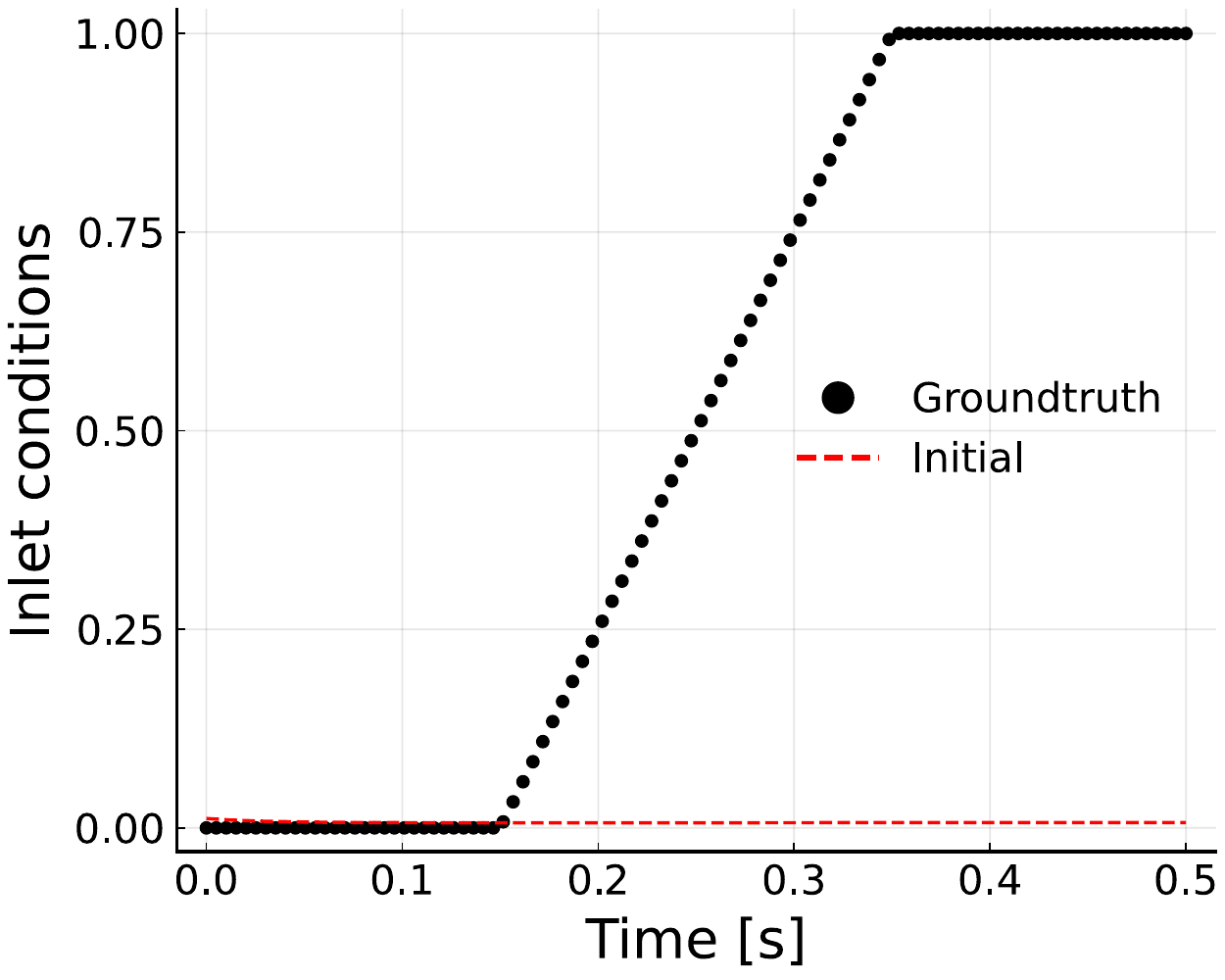}
        \label{figure:fuel_switch_init:a}
    }
    \subfigure[state variables]{
        \includegraphics[width=0.62\textwidth]{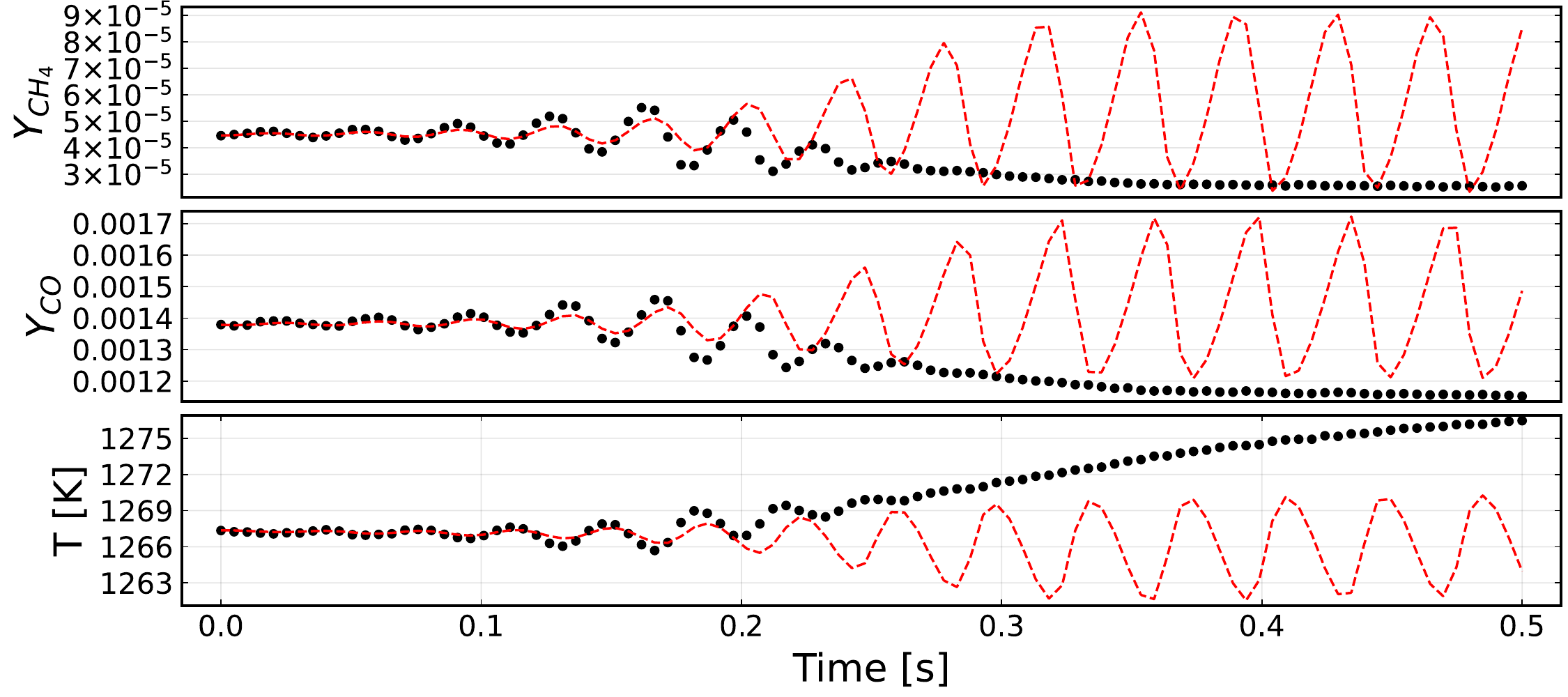}
        \label{figure:fuel_switch_init:b}
    }
    \centering
    \caption{
        Initial states of the fuel switching case. (a) The ground truth and the initialized inlet conditions of COG2 ratio. (b) The ground truth and predicted species and temperature profiles based on the inlet conditions in (a).
    }
\end{figure*}

\begin{figure*}
    \centering
    \subfigure[inlet conditions]{
        \includegraphics[width=0.34\textwidth]{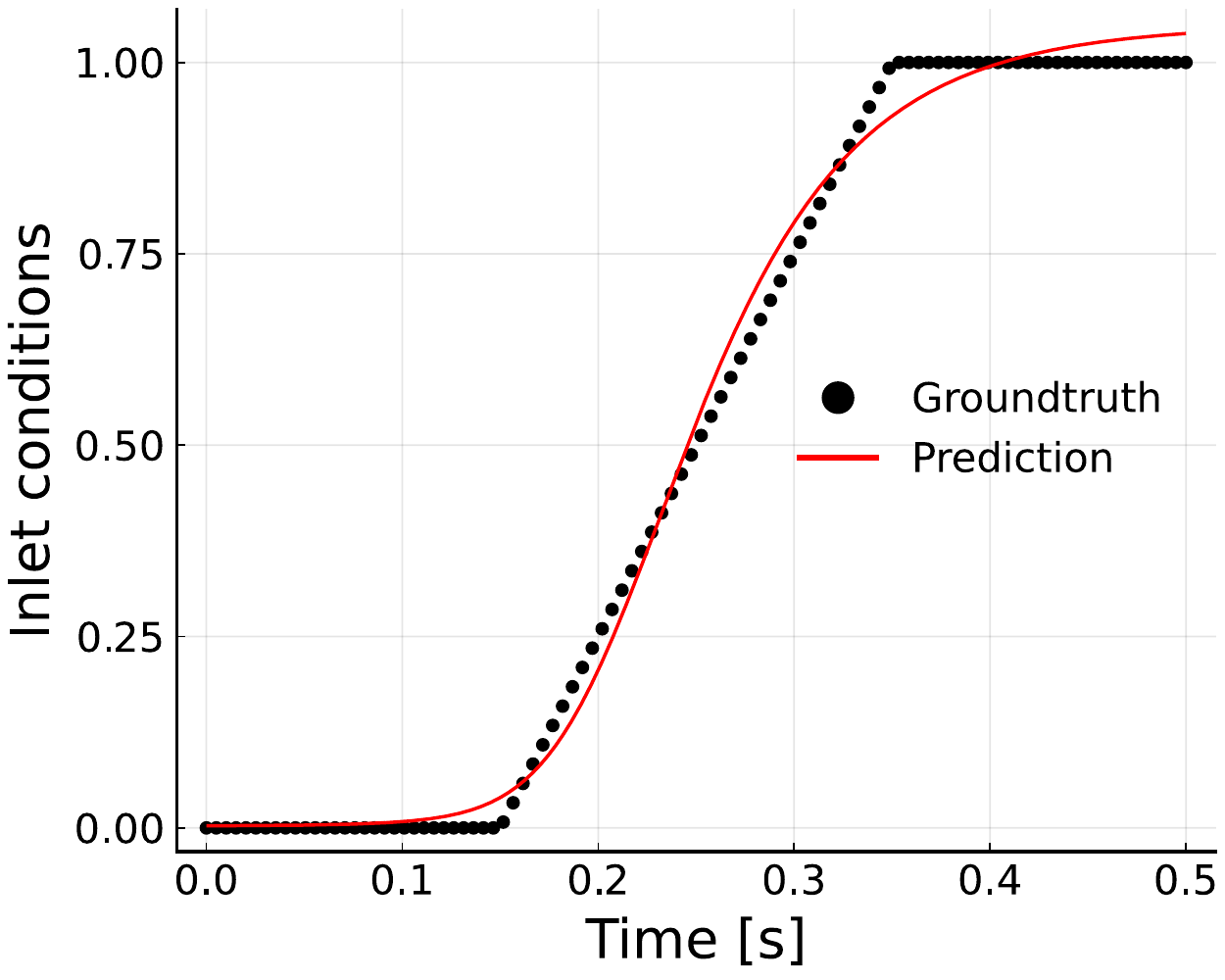}
        \label{figure:fuel_switch_pred:a}
    }
    \subfigure[state variables]{
        \includegraphics[width=0.62\textwidth]{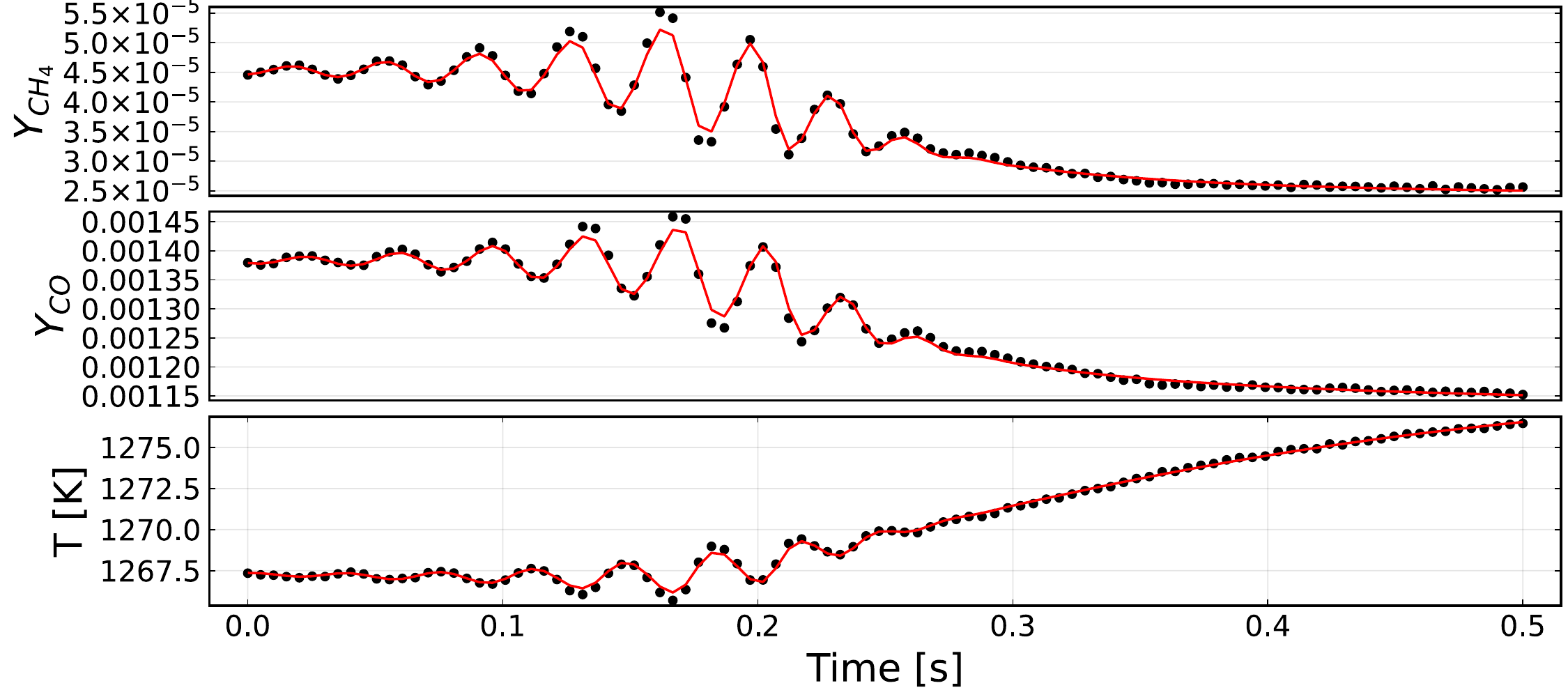}
        \label{figure:fuel_switch_pred:b}
    }
    \centering
    \caption{
        Trained results of the fuel switching case. (a) The ground truth and the inferred inlet conditions of COG2. (b) The ground truth and predicted species and temperature profiles based on the inlet conditions in (a).
    }
\end{figure*}

In combustion devices such as hot air heaters, coke oven gas can be directly ignited at a normal temperature, for it has half of the calorific value of natural gas, about 4.75 $kWh/m^3$. On the contrary, blast furnace gas with one-tenth the calorific value of natural gas, about 0.95 $kWh/m^3$, needs to be burnt at a higher temperature. During the fuel switch, the distinct combustion characteristic between the two gases can induce oscillating combustion, and thus cause the vibration of the heater body and the instability of the heated air. In addition, extinction occasionally occurs since the switching temperature is usually determined based on empirical rules and thus it is not readily known and may not be optimal during the fuel-switching stage. Therefore, identifying the transition process during fuel switching would be beneficial for further optimizing the operation and avoiding unexpected extinction.

Here, we simulate the fuel switching from COG1 to COG2 with a linear switching function to synthesize the measurement dataset. As shown in Fig. \ref{figure:fuel_switch_init:a}, the volume fraction of COG2 at the inlet transitions from 0 to 1 to represent the fuel switching process from COG1 to COG2 in an industrial combustor. As detailed in \cite{zhang2020oscillation}, the burning of COG fuels will show instability from mixing-chemistry interaction, with an oscillation frequency of around 20 Hz. Therefore, the simulation time range here is chosen to be [0, 0.5s] with 100 data points in between, and the fuel switch process occurs between 0.15 s and 0.35 s. Such time interval in the dataset is sufficiently small to resolve the oscillation inside the combustor.

In practice, it is often the case that we can only measure the temperature profiles while it is difficult to measure the species concentrations. Therefore, to simulate this challenging scenario, we only employ the temperature profiles into the loss functions for training the neural network model, i.e.,
\begin{equation} \label{eq:loss_T}
    Loss = MSE\left(T, \widehat{T}\right).
\end{equation}

Since the neural differential equations involve stiff combustion chemistry, differentiating a stiff neural ODE is usually much more computationally expensive than a non-stiff problem, as pointed out in \cite{kim2021stiff}. In addition, the computational cost scales with the number of neural network parameters. To reduce the burden of automatic differentiation, the neural network model should be as concise as possible. Here, a simple neural network model with a single hidden layer with five hidden nodes is employed. The activation functions used are $\tanh(x)$ for the hidden layer and $\exp(x)$ for the output layer.

\begin{figure} [!ht]
    \centering
    \includegraphics[width=1.0\linewidth]{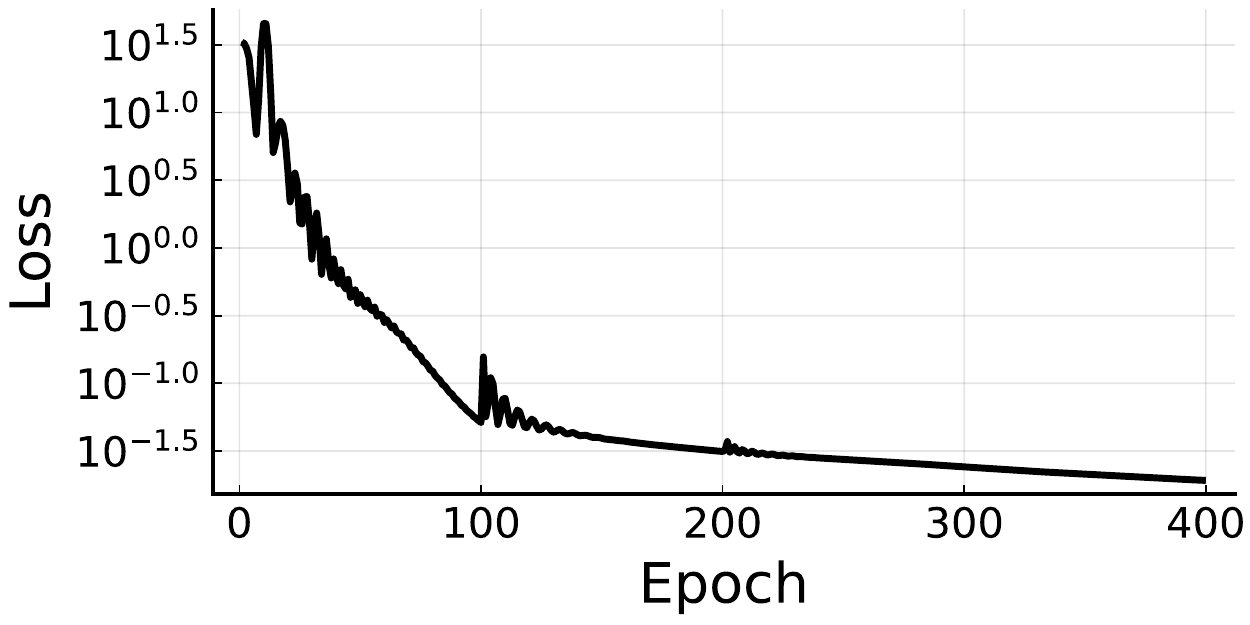}
    \centering
    \caption{The loss history of the fuel switching case.}
    \label{figure:fuel_switch_loss}
\end{figure}

The popular first-order optimizer of Adam \cite{kingma2014adam} is adopted for the optimization. The optimization proceeds step-by-step with a decaying learning rate.
Specifically, we first employ a learning rate of 1e-1 to capture the most important information with 100 training epochs. We then employ 1e-2 for another 100 epochs and 5e-3 for extra 200 epochs to fine-tune the model. Figure \ref{figure:fuel_switch_loss} presents the history of the loss function. As can be seen, the loss decreases stably, which indicates that the learning rate is properly chosen. The loss function reaches a plateau after 300 epochs, and thus 400 epochs are sufficient for the training.

The initialization of the neural network models is achieved by assuming a steady inlet with only COG1. Even though the inlet condition is steady, oscillation is observed in the combustor due to intrinsic instability, as indicated by the dashed line shown in Fig. \ref{figure:fuel_switch_init:b}. Similar oscillation is found in the evolution of the state variables corresponding to the ground truth inlet condition, where fuel switches from COG1 to COG2. Shown as the symbols, 1\% of noise has been added to represent the measurement uncertainty. Such synthesized measurements are also shown in Fig. \ref{figure:fuel_switch_init:b} and utilized to train neural network models to infer the inlet conditions. Therefore, the lines in Figs. \ref{figure:fuel_switch_init:a} and \ref{figure:fuel_switch_init:b} demonstrate the inferred inlet conditions and predicted state variables, respectively. As can be seen, the inferences and predictions agree very well with the ground truth. In addition, the predicted species profiles for methane and carbon monoxide also agree well with the ground truth, although they are not included in the loss function. Such capability of revealing hidden information demonstrates the good generalization capability of the approach. The training takes about 2 hours on a Linux server with a single CPU core, and the training can be potentially accelerated by improving the weight initialization strategy and parallel computing \cite{kim2021stiff}.

\subsection*{3.2 Inverse Modeling for Inflow Fluctuations}

\begin{figure*}
    \centering
    \subfigure[inlet conditions]{
        \includegraphics[width=0.34\textwidth]{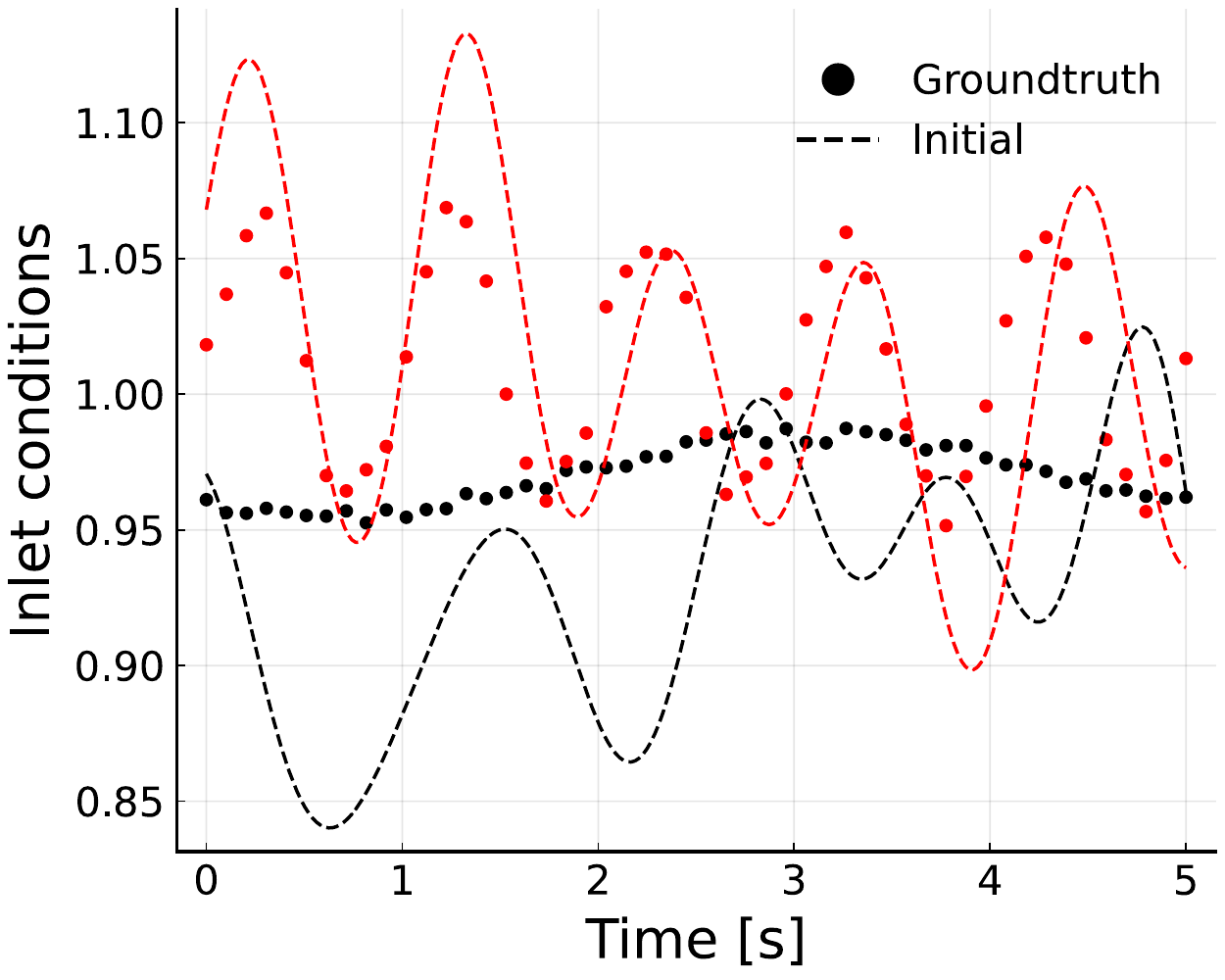}
        \label{figure:fluctuate_noise_init:a}
    }
    \subfigure[state variables]{
        \includegraphics[width=0.62\textwidth]{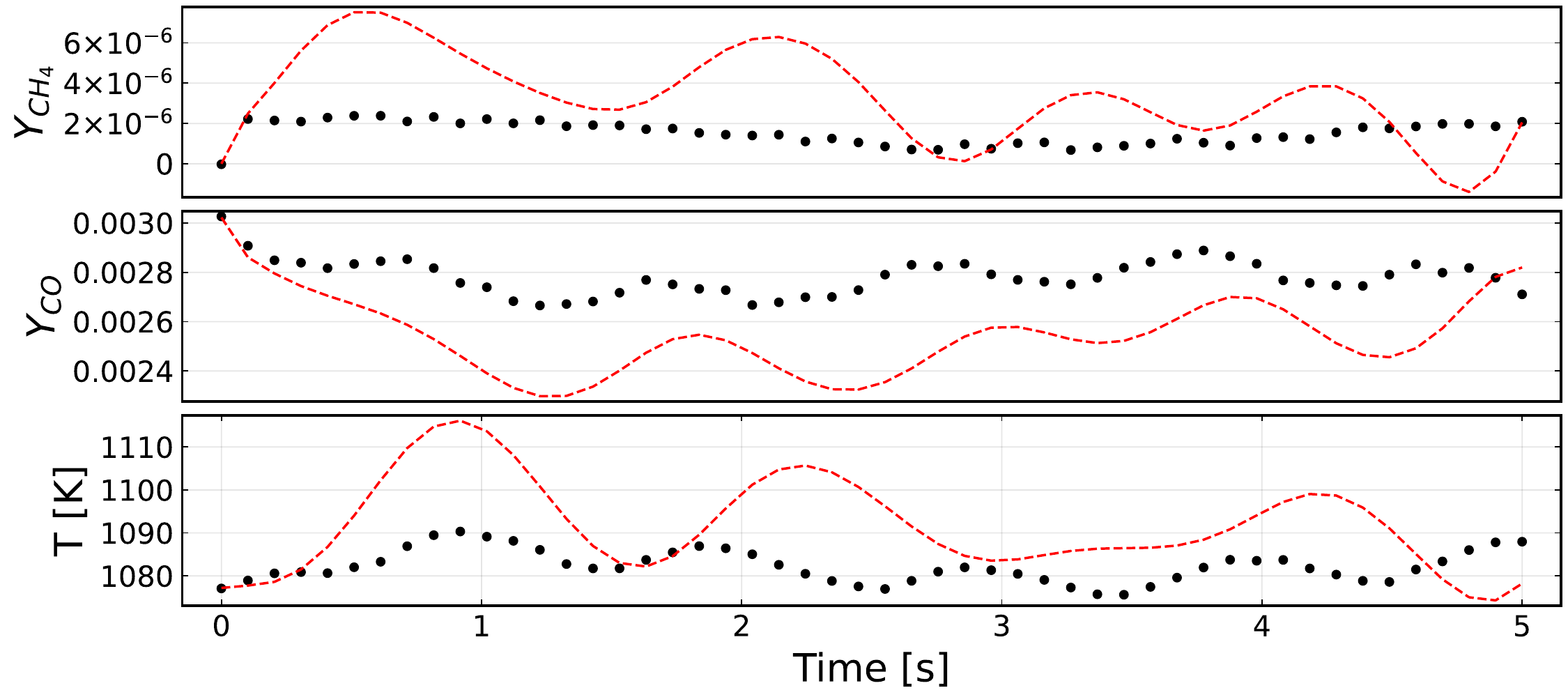}
        \label{figure:fluctuate_noise_init:b}
    }
    \centering
    \caption{
        Initial states of the inflow fluctuating case. (a) The ground truth and the initialized inlet conditions of the volume fraction of BFG (black) and the flow rate represented by $\tau_{res}$ (red). (b) The ground truth and predicted species and temperature profiles based on the inlet conditions in (a).
    }
\end{figure*}

Another important mechanism for oscillating combustion behaviors is the uncertain inlet conditions with flow and calorific fluctuations during the stable burning of the BFG gas. According to Zhang et al., \cite{zhang2021ann}, typical inflow fluctuations of hot air heater consist of $10\%$ of fluctuation in inlet flow rates and $3\%$ of fluctuation in inlet calorific value, with frequencies around 1 Hz and 0.05 Hz, respectively. In this study, we employ a sine function with $20\%$ of noise added to synthesize the "ground truth" inlet boundary conditions, as shown in Fig. \ref{figure:fluctuate_noise_init:a}. The inlet flow rate fluctuation is characterized by the fluctuations in the resident time $ \tau_{res} = \rho V / m_a $, which has a nominal value of $\bar \tau_{res} = 1.0 s$ and frequency of 1 Hz. For simplicity, we assume that the fluctuation of inlet calorific value is due to the composition variation in the fuel blend of COG1 and BFG. Thus, the volume fraction of BFG can well characterize the change in the inlet calorific value. To reduce the computational cost, the frequency of calorific fluctuation is set to be 0.2 Hz and the time range for the ODE integration is [0, 5 s] for demonstration. In addition, to demonstrate the robustness and capability of the feature learning of the current method, 10\% of noise is added to the state variables obtained at the presumed inlet boundary conditions to be utilized as the training dataset for the neural network.

The structure of the neural network $NN_{inlet}$ is similar to that mentioned in the fuel switching case, but has two output neurons corresponding to the two inlet variables in this case, i.e., the volume fraction of BFG and the residence time $\tau_{res}$. Besides, to capture the periodic inlet boundary conditions, we employ $\sin(x)$ as the activation function for the hidden layer. This means that the neural network inlet $NN_{inlet}$ is a linear combination of sine functions
\begin{equation}
    NN_{inlet}(t)_j = \exp\left( \sum_i \left( A_{ij} \sin(a_i x + b_i) + B_{ij} \right) \right),
\end{equation}
where $i=1,2,...,5$, $j=1,2$, correspond to the indices of neurons in the hidden layer and the output layer. $a_i$, $b_i$ are the weights and biases of the mapping from the input layer to the hidden layer. $A_{ij}$ and $B_{ij}$ are the weights and biases of the mapping from the hidden layer to the output layer.

In practice, the fluctuation frequency and amplitude in the temperature profile can provide insights for the initialization of the neural network. For example, Fig.~\ref{figure:fluctuate_noise_init:a} shows the $NN_{inlet}$ initialized with similar frequencies and amplitudes as the temperature profiles, and the corresponding state variables are shown in Fig.~\ref{figure:fluctuate_noise_init:b} with the dashed red lines.

However, due to the complexity of the coupling between flow and calorific fluctuations, it is difficult to identify the boundary conditions with only temperatures as the observed data. Therefore, temperature along with CO mass fractions are used to constrain the neural networks via the loss function 
\begin{equation}
    Loss = MSE\left([T,Y_{CO}], [\widehat{T},\widehat{Y}_{CO}]\right).
\end{equation}

\begin{figure*}
    \centering
    \subfigure[inlet conditions]{
        \includegraphics[width=0.34\textwidth]{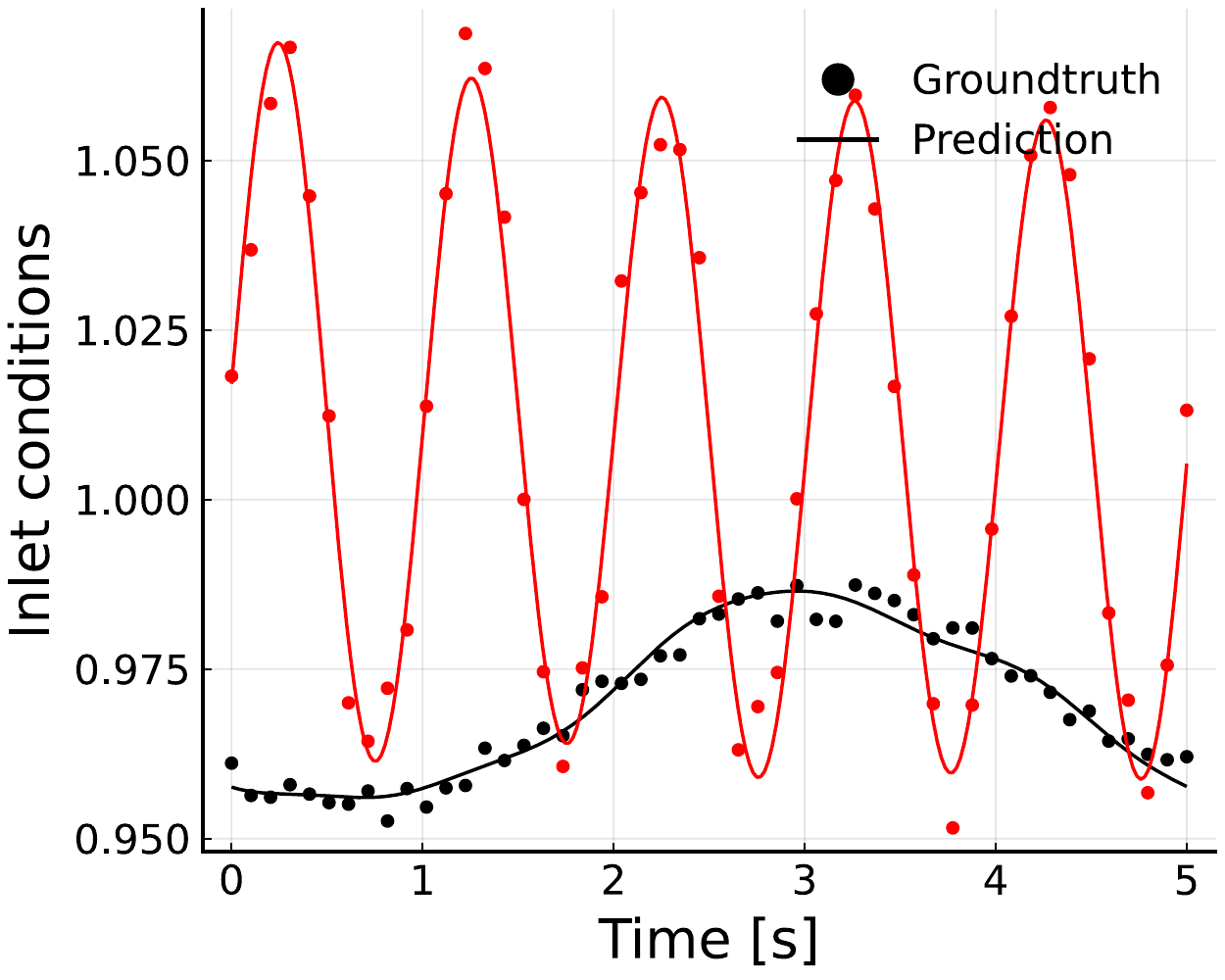}
        \label{figure:fluctuate_noise_pred:a}
    }
    \subfigure[state variables]{
        \includegraphics[width=0.62\textwidth]{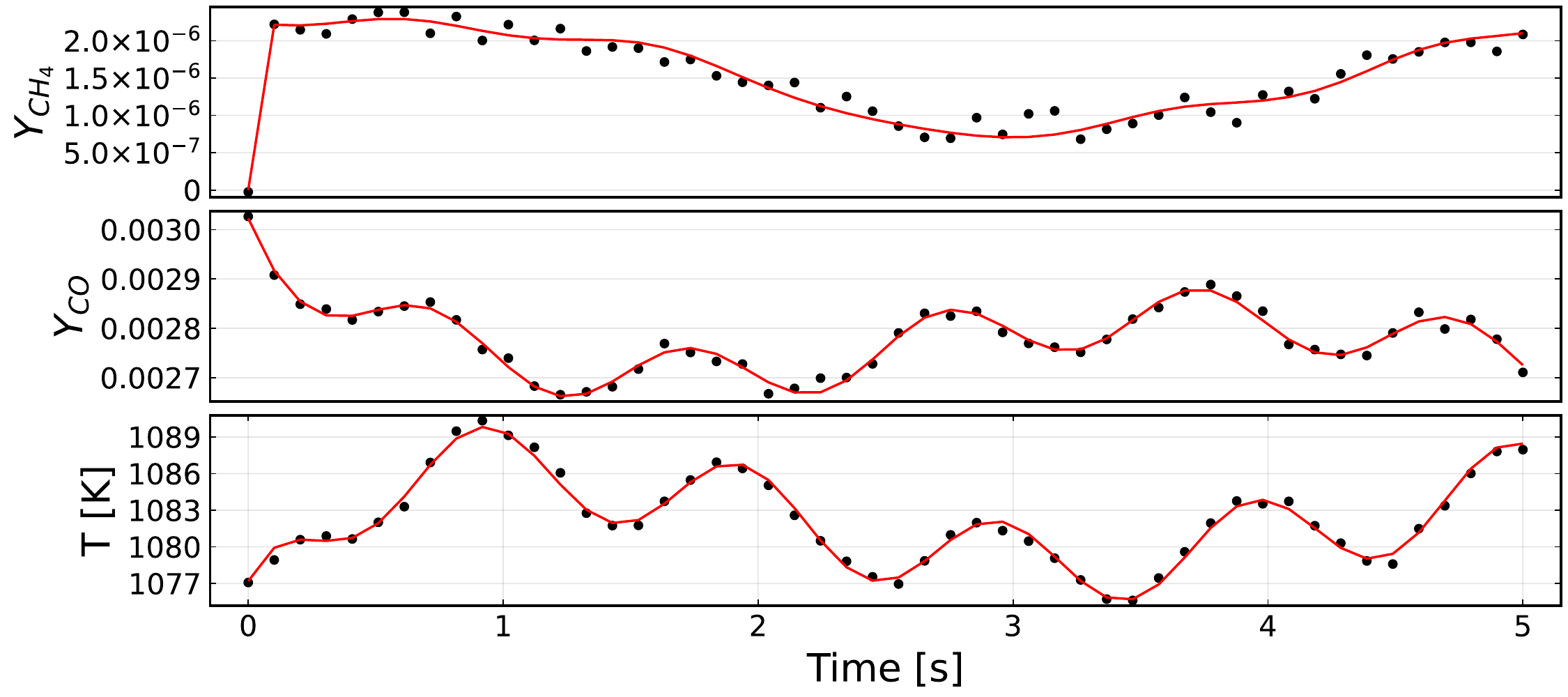}
        \label{figure:fluctuate_noise_pred:b}
    }
    \centering
    \caption{
        Trained results of the inflow fluctuating case. (a) The ground truth and the inferred inlet conditions of the volume fraction of BFG (black) and the flow rate represented by $\tau_{res}$ (red). (b) The ground truth and predicted species and temperature profiles based on the inlet conditions in (a).
    }
    \label{figure:fluctuate_noise_pred}
\end{figure*}

\begin{figure} [!ht]
    \begin{center}
    \includegraphics[width=1.0\linewidth]{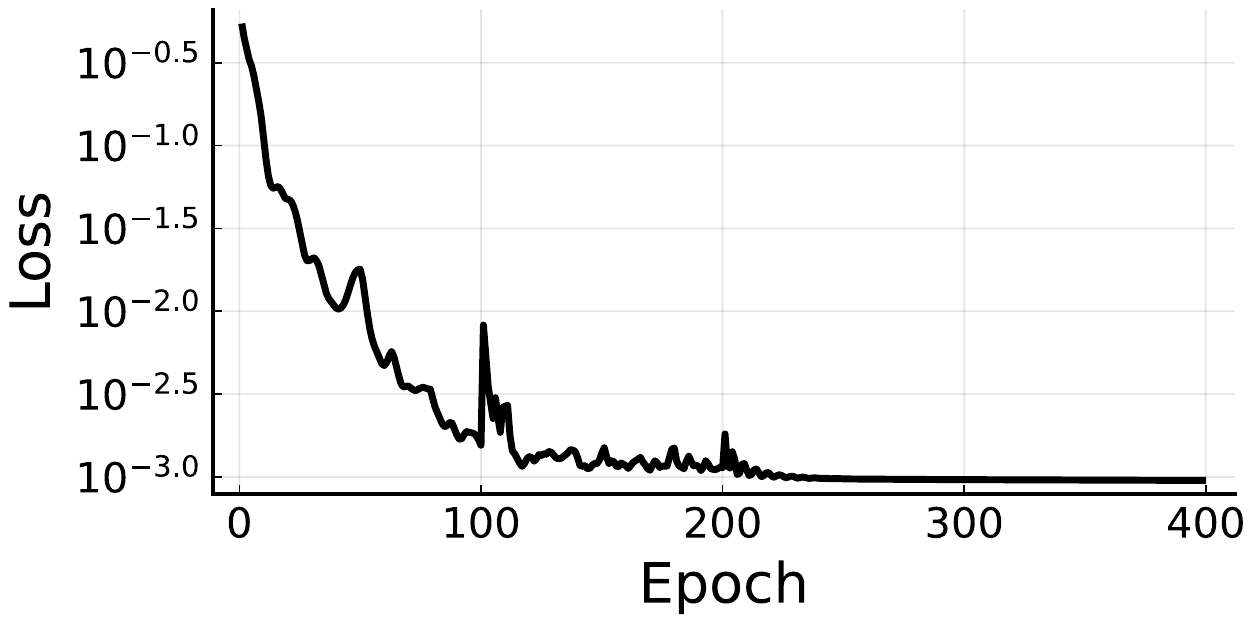}
    \end{center}
    \caption{The loss history of the inflow fluctuating case.}
    \label{figure:fluctuate_noise_loss}
\end{figure}

Similar to the case of fuel switching, we gradually decaying the learning rate. We first employ a learning rate of 5e-3 to capture the most important information with 100 training epochs. Then a learning rate of 2e-3 for another 100 epochs and 1e-3 for extra 200 epochs are employed to fine-tune the results. The training results are shown in Fig.~\ref{figure:fluctuate_noise_pred}, which takes about 30 minutes to train with a single thread. The losses of the training process are shown in Fig.~\ref{figure:fluctuate_noise_loss}.

As can be seen from Figs.~\ref{figure:fluctuate_noise_pred:a} and \ref{figure:fluctuate_noise_pred:b}, the inferred inlet conditions and the predicted temperature evolution with the inferred inlet conditions agree with the ground truth very well. In addition, the predicted species profiles for methane and other species also agree well with the ground truth. Such capability of revealing coupled hidden inlet conditions further shows the potential of the approach in fault diagnostics for sourcing the causes of combustion oscillation.

\section*{4. CONCLUSION} \label{sec:conclusion}

This work presents a machine learning approach for inverse modeling of inlet boundary conditions of a model combustor. The approach is demonstrated to recover the inlet conditions under two common modes of unsteady inlets: fuel switching and inlet flow rate/calorific fluctuations. Using synthesized data, the results show that the approach can accurately infer the temporal evolution of the inlet boundary conditions from limited measurements of temperature and/or CO profiles. In addition, the temporal profiles of methane inside the combustor can also be accurately inferred although methane is not exposed to the training process. Therefore, the approach is shown to efficiently and accurately reveal the hidden dynamics of the upstream compositions and flow rate.

The major limitations of this approach are that the algorithm relies on an accurate combustion kinetic mode, and there could be multiple possible solutions to the inference due to the uncertainties in the experimental data. In future works, we will incorporate Bayesian neural networks \cite{gal2016dropout} to quantify impact of the uncertainties in the model and the experimental data to the inferences solution. The estimated uncertainty will greatly facilitate the development of dynamic control of industrial combustion systems.

\begin{acknowledgment}

ZR and XS would like to acknowledge the support from National Natural Science Foundation of China No. 52025062.

\end{acknowledgment} 

\bibliographystyle{asmems4}
\bibliography{ms}

\end{document}